\documentclass[11pt]{article}
\usepackage{longtable,supertabular}
\usepackage{amsmath}
\usepackage{amssymb}
\usepackage{latexsym}
\usepackage{cite}

\title{\bf MDS Entanglement-Assisted Quantum Codes of Arbitrary Lengths and Arbitrary Distances}
\author{Hao Chen
  \thanks{Hao Chen is with the College of Information Science and Technology/Cyber Security, Jinan University, Guangzhou, Guangdong Province, 510632, China, haochen@jnu.edu.cn. The research of Hao Chen was supported by NSFC Grant 62032009.}}

\begin{document}

\maketitle
\begin{abstract}
Quantum error correction is fundamentally important for quantum information processing and computation. Quantum error correction codes have been studied and constructed since the pioneering papers of Shor and Steane. Optimal (called MDS) $q$-qubit quantum codes attaining the quantum Singleton bound were constructed for very restricted lengths $n \leq q^2+1$. Entanglement-assisted quantum error correction (EAQEC) code was proposed to use the pre-shared maximally entangled state for the enhancing of error correction capability. Recently there have been a lot of constructions of MDS EAQEC codes attaining the quantum Singleton bound for very restricted lengths. In this paper we construct such MDS EAQEC $[[n, k, d, c]]_q$ codes for arbitrary $n$ satisfying $n \leq q^2+1$ and arbitrary distance $d\leq \frac{n+2}{2}$. It is proved that for any given length $n$ satisfying $O(q^2)=n \leq q^2+1$ and any given distance $d$ satisfying $ O(q^2)=d \leq \frac{n+2}{2}$, there exist  at least $O(q^2)$ MDS EAQEC $[[n, k, d, c]]_q$ codes with different $c$ parameters. Our results show that there are much more MDS entanglement-assisted quantum codes than MDS quantum codes without consumption of the maximally entangled state. This is natural from the physical point of view. Our method can also be applied to construct MDS entanglement-assisted quantum codes from the generalized MDS twisted Reed-Solomon codes.\\

{\bf Index terms:} MDS quantum code, MDS entanglement-assisted quantum code, Hermitian hull.
\end{abstract}

\section{Introduction}

Quantum error correction is fundamentally important for quantum information processing and quantum computation. The Calderbank-Shor-Steane stabilizer construction of quantum error correction codes from classical error-correcting codes was proposed in \cite{Shor,Steane,CShor,Steane1}. Then asymptotically good quantum error correction codes and many quantum codes from classical codes were constructed, see \cite{CRSS,HChen1,HChen2,HChen3,AKS,KS,Ball1}. In \cite{Brun,Fattal} entanglement-assisted quantum error correction (EAQEC) code was proposed and the similar CSS stabilizer construction was given. Comparing to an QECC, an EAQEC code has one more parameter $c$ measuring the consumption of $c$ pre-shared copies of the maximally  entangled state. We refer to \cite{BDH,HDB,HBD,Barta,LBW,WKB,KHB,WB,Fujiwara} for related development of entanglement-assisted quantum error correction. There have been a lot of effort to construct MDS entanglement-assisted quantum codes with specific parameters attaining the quantum Singleton bound, for example, see \cite{Kor,LCC,LES22,LEGL22,GYHZ,FFLZ20,GMCR,CMPR,CZJL,CZJ} and references therein. One basic problem in entanglement-assisted quantum error correction is that there are how many such optimal codes.\\

The quantum Singleton bound for an $[[n, k, d]]_q$ quantum code is $$d+2k \leq n+2.$$ On the other hand the quantum Singleton bound claims $$2d+k\leq n+c+2$$ for an EAQEC $[[n, k, d, c]]_q$ code when $d\leq \frac{n+2}{2}$, see \cite{Brun,GHW}. Therefore an EAQEC code could have larger $k$ and $d$ because of the consumption of pre-shared $c$ copies of the maximally entangled state. A quantum code satisfying $d+2k=n+2$ is called an MDS quantum code. An EAQEC code satisfying $2d+k=n+c+2$ and $d \leq \frac{n+2}{2}$  is called an MDS entanglement-assisted quantum code. MDS quantum codes and MDS entanglement-assisted quantum codes are considered as optimal quantum codes with the best possible parameters. The defect $D({\bf C})$ of an EAQEC $[[n, k, d, c]]_q$ code ${\bf C}$ is defined by $$D({\bf C})=(n+c+2)-(2k+d)$$ to measure its difference to the quantum Singleton bound.  MDS EAQEC codes constructed in papers \cite{LCC,FFLZ20,GYHZ,LES22} are these codes with $D({\bf C})=0$. Almost MDS EAQEC codes constructed in \cite{Pellikaan,HChen22} are the EAQEC codes satisfying $D({\bf C})=2$.  Quantum error correction codes from BCH codes were constructed in \cite{AKS}. The defects of these quantum BCH codes can be determined explicitly and some of them are small when compared to the lengths. The construction of linear codes with $h$-dimension Hermitian hull and their applications in EAQEC codes have been an active topic in recent years, we refer to \cite{LCC,GYHZ,Pellikaan}. Entanglement-assisted concatenated codes was proposed in a recent paper \cite{FanPoor}.\\

There are few Hermitian self-dual MDS codes, while there have been a lot of Hermitian self-orthogonal (or dual-containing) codes to construct MDS quantum codes, see \cite{Guar,KZ13,KZL14,HCX,Ball1,JX} and references therein. However these MDS quantum codes were only constructed for some special lengths. We refer to \cite{Ball1,Ball2,Ball3} for recent works on Hermitian self-orthogonal codes and \cite{Ball4} for a nice survey on quantum codes. It is still unknown if  MDS quantum codes exist for every length $n \leq q^2+1$. To conjecture that there are much more MDS entanglement-assisted quantum codes than MDS quantum codes seems reasonable from the physical point of view. In our previous paper we proved that MDS entanglement-assisted quantum $[n, k, d, c]]_q$ codes with nonzero $c$ parameters exist for any given length $n \leq q^2+1$. \\

In this paper we prove that for any given length $n$ satisfying $O(q^2)=n \leq q^2+1$ and any given distance $O(q^2)=d \leq \frac{n+2}{2}$, there exist at least $O(q^2)$ MDS EAQEC $[[n, k, d, c]]_q$ codes with different $c$ parameters. Since MDS quantum codes have been only constructed for some restricted lengths. Our result shows that there are much more MDS entanglement-assisted quantum codes than MDS quantum codes. Our method can also be applied to give a lower bound of the dimensions of Hermitian hulls of a class of generalized twisted Reed-Solomon codes. Then MDS entanglement-assisted quantum codes are also constructed from the generalized MDS twisted Reed-Solomon codes.\\

\section{Preliminaries}

The Hamming weight of a vector ${\bf a} \in {\bf F}_q^n$ is the number of non-zero coordinate positions. The Hamming distance $d({\bf a}, {\bf b})$ between two vectors ${\bf a}$ and ${\bf b}$ is the Hamming weight of ${\bf a}-{\bf b}$. For a code ${\bf C} \subset {\bf F}_q^n$, its minimum Hamming distance $$d({\bf C})=\min_{{\bf a} \neq {\bf b}} \{d({\bf a}, {\bf b}),  {\bf a} \in {\bf C}, {\bf b} \in {\bf C} \},$$  is the minimum of Hamming distances $d({\bf a}, {\bf b})$ between any two different codewords ${\bf a}$ and ${\bf b}$ in ${\bf C}$. The shortening code of ${\bf C}$ at the $i$-th coordinate position is the subcode of ${\bf C}$ consisting of codewords in ${\bf C}$ whose $i$-th coordinates are zero. The punctured code of ${\bf C}$ at the $i$-th position is the image of ${\bf C}$ under the natural projection from ${\bf F}_q^n$ to ${\bf F}_q^{n-1}$ by deleting the $i$-th coordinate. The minimum Hamming distance of  a linear code is its minimum Hamming weight.  The Singleton bound asserts $d \leq n-k+1$ for a linear $[n, k, d]_q$ code. When the equality holds, this code is an MDS code. The main conjecture of MDS codes claims that the length of an MDS code over ${\bf F}_q$ is at most $q+1$, except some trivial exceptional cases. In \cite{Ball} the main conjecture of MDS codes was proved for codes over prime fields. We refer to \cite{HP} for the theory of Hamming error-correcting codes.\\

The Euclidean inner product on ${\bf F}_q^n$ is defined by $$<{\bf x}, {\bf y}>=\Sigma_{i=1}^n x_iy_i,$$ where ${\bf x}=(x_1, \ldots, x_n)$ and ${\bf y}=(y_1, \ldots, y_n)$. The Euclidean dual of a linear code ${\bf C}\subset {\bf F}_q^n$ is $${\bf C}^{\perp}=\{{\bf c} \in {\bf F}_q^n: <{\bf c}, {\bf y}>=0, \forall {\bf y} \in {\bf C}\}.$$ The Hermitian inner product on ${\bf F}_{q^2}^2$ is defined by  $$<{\bf x}, {\bf y}>_H=\Sigma_{i=1}^n x_iy_i^q,$$ where ${\bf x}=(x_1, \ldots, x_n)$ and ${\bf y}=(y_1, \ldots, y_n)$ are two vectors in ${\bf F}_{q^2}^n$. The Hermitian dual of a linear code ${\bf C} \subset {\bf F}_{q^2}^n$ is $${\bf C}^{\perp_H}=\{{\bf c} \in {\bf F}_{q^2}^n: <{\bf c}, {\bf y}>_H=0, \forall {\bf y} \in {\bf C}\}.$$  It is clear ${\bf C}^{\perp_H}=({\bf C}^{\perp})^q$, where $${\bf C}^q=\{(c_1^q, \ldots, c_n^q): (c_1 \ldots, c_n) \in {\bf C}\}.$$ The minimum distance of the Euclidean dual is called the dual distance and is denoted by $d^{\perp}$. The minimum distance of the Hermitian dual is the same as $d^{\perp}$. A linear code ${\bf C} \subset {\bf F}_{q^2}^n$ is called Hermitian self-dual if ${\bf C}={\bf C}^{\perp_H}$, Hermitian self-orthogonal if ${\bf C} \subset {\bf C}^{\perp_H}$, and Hermitian dual-containing if ${\bf C}^{\perp_H} \subset {\bf C}$. The linear code ${\bf C} \bigcap {\bf C}^{\perp}$ is called the Euclidean hull of the linear code ${\bf C} \subset {\bf F}_q$. The intersection ${\bf C} \bigcap {\bf C}^{\perp_H}$ is called the Hermitian hull of the linear code ${\bf C} \subset {\bf F}_{q^2}^n$. \\

Two codes ${\bf C}_1$ and ${\bf C}_2$ in ${\bf F}_q^n$ are equivalent if and only if ${\bf C}_2$ can be obtained from ${\bf C}_1$ by a permutation of coordinates and the multiplication of a Hamming weight $n$ vector ${\bf v}=(v_1, v_2, \ldots, v_n) \in {\bf F}_q^n$ on coordinates, where $v_i \neq 0$ for $i=1, \ldots, n$. That is $${\bf C}_2=\{{\bf c}=(c_1, \ldots c_n): (c_1, \ldots, c_n)=(v_1x_1, \ldots, v_nx_n), {\bf x} \in Perm({\bf C}_1)\},$$ where $Perm({\bf C}_1)$ is the code obtained from ${\bf C}_1$ by a coordinate permutation. Equivalent codes have the same distances and weight distributions.  Obviously equivalent codes have different dual codes as follows. Let ${\bf v}$ be a Hamming weight $n$ vector ${\bf v}=(v_1, v_2, \ldots, v_n) \in {\bf F}_q^n$, set ${\bf v} \cdot {\bf C}=\{(v_1c_1, \ldots, v_nc_n): \forall {\bf c}=(c_1, \ldots, c_n) \in {\bf C}\}$. Then $$({\bf v} \cdot {\bf C})^{\perp}={\bf v}^{-1} \cdot {\bf C}^{\perp},$$ where ${\bf v}^{-1}=(v_1^{-1}, \ldots, v_n^{-1}).$ For a linear code ${\bf C} \subset {\bf F}_{q^2}^n$ and a Hamming weight $n$ vector ${\bf v} \in {\bf F}_{q^2}^n$, then $$({\bf v} \cdot {\bf C})^{\perp_h}={\bf v}^{-q} \cdot {\bf C}^{\perp_h},$$ where ${\bf v}^{-q}=(v_1^{-q}, \ldots, v_n^{-q})$.\\

The CSS stabilizer construction of quantum code implies that a quantum $[[n, 2k-n, d]]_q$ code can be obtained from a Hermitian dual-containing $[n, k, d]_{q^2}$ code and a quantum $[[n, n-2k, d]]_q$ code can be obtained from a Hermitian self-orthogonal $[n, k, d]_{q^2}$ code. From the CSS construction of entanglement-assisted quantum code in \cite{Brun}, we recall the following result in \cite{Brun}.\\

{\bf CSS construction of entanglement-assisted quantum codes.} {\em Let ${\bf C} \subset {\bf F}_{q^2}$ be a linear $[n, k, d]_{q^2}$ code with the $h$-dimensional Hermitian hull. Then an EAQEC $[[n, k-h, d, n-k-h]]_q$ code and an EAQEC $[[n, n-k-h, d^{\perp}, k-h]]_q$ code can be constructed. When $d \leq \frac{n+2}{2}$ or $d^{\perp} \leq \frac{n+2}{2}$ and the code is MDS, then the above code is an MDS EAQEC code.}\\

In the above construction of EAQEC codes, the dimension $h$ of the hull of a linear code is the key parameter to control the dimension and consumption parameter of an EAQEC code, since the other parameters of codes are preserved when a linear code is transformed to an equivalent linear code. This is an important motivation to construct equivalent linear codes with various dimension Hermitian hulls.  In \cite{HChen22} it was proved for a linear $[n, k, d]_{q^2}$ code ${\bf C} \subset {\bf F}_{q^2}^n$ with the Hermitian hull of the dimension $\dim(Hermitianhull) \geq h$, then there exist at least  $h+1$ different entanglement-assisted quantum codes with different $c$ parameters and the same defect. On the other hand from the CSS construction of entanglement-assisted quantum codes, if the dimension of the Hermitian hull is large, a better EAQEC code with the smaller $c$ parameter and the same defect can be constructed. From this motivation it is natural to ask that for fixed $[n, k, d]_{q^2}$ linear codes, what is the largest possible Hermitian hull? We refer to \cite{HChen22} for the study of this problem. In this paper we give a lower bound of the dimensions of Hermitian hulls of a class of generalized twisted Reed-Solomon codes.\\

\section{Arbitrary length MDS codes with lower bounded dimension Hermitian hulls}

We need the following result to adjust the dimensions of Hermitian hulls to construct entanglement-assisted quantum codes.\\

{\bf Proposition 3.1.} {\em Let $q$ be a prime power satisfying $q \geq 3$. Let ${\bf C} \subset {\bf F}_{q^2}$ be a linear $[n, k]_{q^2}$ code with $l$-dimension Hermitian hull. Then for any nonnegative integer $l' \leq l$ there is a Hamming weight $n$ vector ${\bf v}$ such that ${\bf v} \cdot {\bf C}$ has the $l' $ dimension Hermitian hull. Thus from an arbitrary linear $[n, k]_{q^2}$ code ${\bf C} \subset {\bf F}_{q^2}^n$ with its dual distance $d^{\perp} \leq \frac{n+2}{2}$ and the $h$-dimension Hermitian hull, we have an EAQEC $[[n, n-k-l, d^{\perp}, k-l]]_q$ code, for each nonnegative integer $l$ satisfying  $l \leq h$.}\\

{\bf Proof.} We can assume that the generator matrix of ${\bf C}$ is of the following form,

$$
\left(
\begin{array}{ccccccc}
{\bf I}_l&{\bf 0}_{l, k-l}&{\bf P}\\
{\bf  0}_{k-l, l}&{\bf I}_{k-l}&{\bf Q}\\
\end{array}
\right)
$$
The first $l$ rows is an $l \times n$ matrix, which is a generator matrix of ${\bf C} \bigcap {\bf C}^{\perp_H}$. Then ${\bf P} \cdot {\bf \bar{P}}^{\tau}=-{\bf I}_l$. Set ${\bf Q}=({\bf Q}_1, {\bf Q}_2)$, where ${\bf Q}_1$ is a $(k-l) \times (k-l)$ matrix and ${\bf Q}_2$ is a $(k-l) \times (n-k-l)$ matrix. By multiplying a Hamming weight $n$ vector ${\bf v}=(\lambda_1, \lambda_{l-l'}, 1, \ldots, 1) \in {\bf F}_{q^2}^n$ such that $\lambda_i^{q+1}$ is not $1$. Now we prove that  the Hermitian hull is just the subcode generated by $l-l'+1, \ldots, l$ rows.\\

Then the generator matrix ${\bf G}_{{\bf v} \cdot {\bf C}}$ of ${\bf v} \cdot {\bf C}$ is of the following form,\\

$$
\left(
\begin{array}{ccccccc}
{\bf D_{\lambda}}&{\bf 0}_{l, k-l}&{\bf P}_1&{\bf P}_2\\
{\bf  0}_{k-l, l}&{\bf I}_{k-l}&{\bf Q}_1&{\bf Q}_2\\
\end{array}
\right)
$$

where ${\bf D_{\lambda}}$ is a $l \times l$ non-singular matrix of the following form.\\

$$
\left(
\begin{array}{ccccccccccccc}
\lambda_1&0&0&\cdots&\cdots&\cdots&\cdots&0\\
0&\lambda_2&0&\cdots&\cdots&\cdots&\cdots&0\\
\cdots&\cdots&\cdots&\cdots&\cdots&\cdots&\cdots&\cdots\\
0&0&0&\cdots&\lambda_{l-l'}&0&\cdots&0\\
0&0&0&\cdots&0&1&\cdots&0\\
0&0&0&\cdots&0&0&\cdots&1\\
\end{array}
\right)
$$
The Hermitian dual ${\bf C}^{\perp_H}$ has one generator matrix ${\bf G}_{{\bf C}^{\perp_H}}$ of the form $(-{\bf \bar{P}}^{\tau}, -{\bf \bar{Q}}^{\tau}, {\bf I}_{n-k})$. Consider the following $(n-k) \times (n-k)$ nonsingular matrix ${\bf W}$,

$$
\left(
\begin{array}{cccc}
{\bf P}_1&{\bf P}_2\\
{\bf 0}_{n-k-l, l}&{\bf I}_{n-k-l}\\
\end{array}
\right)
$$

Then ${\bf W} \cdot {\bf G}_{{\bf C}^{\perp_H}}$ of the following form is also one generator matrix of ${\bf C}^{\perp_H}$. Here ${\bf 0}_{l, k-l}$ and ${\bf 0}_{n-k-l, l}$ are the $l \times (k-l)$ and $(n-k-l) \times l$ zero matrix.\\

$$
\left(
\begin{array}{cccc}
{\bf I}_l&{\bf 0}_{l, k-l}&{\bf P}_1&{\bf P}_2\\
-{\bf \bar{P_2}}^{\tau}&-{\bf \bar{Q_2}}^{\tau}&{\bf 0}_{n-k-l, l}&{\bf I}_{n-k-l}\\
\end{array}
\right)
$$

Hence the Hermitian dual $({\bf v} \cdot {\bf C})^{\perp_H}={\bf v}^{-q} \cdot {\bf C}^{\perp_H}$ has one generator matrix ${\bf G}_{{\bf v}^{-q}  \cdot {\bf C}^{\perp_H}}$ of the  following form.\\

$$
\left(
\begin{array}{cccc}
{\bf D}_{\lambda}^{-q}&{\bf 0}_{l, k-l}&{\bf P}_1&{\bf P}_2\\
-{\bf \bar{P_2}}^{\tau} \dot {\bf D}_{\lambda}^{-q}&-{\bf \bar{Q_2}}^{\tau}&{\bf 0}_{n-k-l, l}&{\bf I}_{n-k-l}\\
\end{array}
\right)
$$

Then last $l'$ rows of the above generator matrix of ${\bf C}$ and the $l-l'+1, l-l'+2, \ldots, l$ rows of the above generator matrix  of $({\bf v} \cdot {\bf C})^{\perp_H}={\bf v}^{-q} \cdot {\bf C}^{\perp_H}$ are the same. Thus the dimension of the Hermitian hull $({\bf v} \cdot {\bf C}) \bigcap ({\bf v} \cdot {\bf C})^{\perp_H}$ is at least $l'$.\\

It is clear that ${\bf C} \bigcap {\bf C}^{\perp_H}$ is the code generated by the first $l$ rows of both generator matrices. It is Hermitian self-orthogonal. From Theorem 2.1 in \cite{HChen221}, ${\bf v} ({\bf C} \bigcap {\bf C}^{\perp_H}) \bigcap {\bf v}^{-q} ({\bf C} \bigcap {\bf C}^{\perp_H})^{\perp_H}$ is of the dimension $l'$. On the other hand $${\bf v}^{-q} \cdot ({\bf C} \bigcap {\bf C}^{\perp_H})^{\perp_H}={\bf v}^{-q}({\bf C}^{\perp_H}\bigcup {\bf C}),$$ has one generator matrix as follows.\\

$$
\left(
\begin{array}{cccc}
{\bf D}_{\lambda}^{-q}&{\bf 0}_{l, k-l}&{\bf P}_1&{\bf P}_2\\
-{\bf \bar{P_2}}^{\tau} \dot {\bf D}_{\lambda}^{-q}&-{\bf \bar{Q_2}}^{\tau}&{\bf 0}_{n-k-l, l}&{\bf I}_{n-k-l}\\
{\bf  0}_{k-l, l}&{\bf I}_{k-l}&{\bf Q}_1&{\bf Q}_2\\
\end{array}
\right)
$$

Then the conclusion follows from Theorem 2.1 in \cite{HChen221} immediately.\\

{\bf Theorem 3.1.} {\em Let $q\geq 3$ be a prime power. For any given length  $n \leq q^2+1$ and any given distance $d \leq \frac{n+2}{2}$, there exists at least one MDS EAQEC $[[n, d, k, c]]_q$ code with nonzero $c$ parameter.}\\

{\bf Proof.} For each length $n \leq q^2+1$ and distance $d \leq \frac{n+2}{2}$ we take a Reed-Solomon $[n, n-d+1, d]_{q^2}$ code ${\bf C}$. Suppose that the dimension of the Hermitian hull of this code is $h$. If $h \geq 1$. Then we have at least $h$ equivalent $[n ,n-d+1, d]_{q^2}$ linear codes with the $l$-dimension Hermitian hull, where $0 \leq l \leq h-1$. Then we have $h$ MDS entanglement-assisted quantum $[[n, n-d+1-l, d, d-1-l]]_q$ codes with nonzero $c$ parameters, since $l \leq n-k-1=d-2$. Even when this Reed-Solomon code is Hermitian LCD, we have at least one MDS entanglement-assisted quantum $[[n, n-d+1, d, d-1]]_q$ code with the nonzero $c$ parameter.\\

From the above proof for any given length $n \leq q^2+1$ and given distance $d\leq \frac{n+2}{2}$ it is good to have a generalized Reed-Solomon $[n, n-d+1, d]_{q^2}$ code with a $h$-dimension Hermitian hull such that $h$ is relatively large. Then we have at least $h$ different MDS entanglement-assisted quantum $[[n, n-d+1-l, d, d-1-l]]_q$ codes, for $l=0, \ldots, h-1$. In the following part of this section we construct many such generalized Reed-Solomon codes for many lengths and many distances such that $h$ can have a good lower bound. In general it is good to construct linear codes over ${\bf C}_{q^2}$ with large Hermitian hulls.\\

{\bf Theorem 3.2.} {\em Let $n$ be a positive integer satisfying $n| q^2-1$ and $k$ be any given positive integer satisfying $k \geq \frac{n}{2}$. Then there exists an MDS linear code ${\bf C} \subset {\bf F}_{q^2}^n$ such that $\dim({\bf C} \bigcap {\bf C}^{\perp_H}) \geq \frac{k(n-k-2)}{q^2}$.}\\

{\bf Proof.} Let ${\bf G}=\{a_1, \ldots, a_n\}$ be all elements of the subgroup of the multiplicative group ${\bf F}_{q^2}^*$ of the order $n$. They are $n$ distinct roots of $$h(x)=x^n-1=\prod_{i=1}^n (x-a_i).$$ Set $RS(n, k)=\{(f(a_1), \ldots, f(a_n)): f \in {\bf F}_{q^2}[x], \deg(f) \leq k-1\}$. Then from the computation in \cite{JX17}, $RS(n, k)^{\perp}={\bf U}^{-1} \cdot RS(n, n-k)$, where ${\bf U}=(u_1, \ldots, u_n) \in ({\bf F}_{q^2}^*)^n$, and $$u_i=(a_i-a_1)\cdots(a_i-a_{i-1})(a_i-a_{i+1})\cdots (a_i-a_n).$$ Thus $$({\bf U}^{-1} \cdot RS)^{\perp}={\bf U}^{-1} \cdot RS(n, k)^{\perp}=RS(n, n-k),$$ and $${\bf U}^{-1} \cdot RS(n, k)=\{(\frac{xf}{n}(a_1), \ldots, \frac{xf}{n}(a_n)): \deg (f) \leq k-1\},$$ where $\frac{xf}{n}$ is the polynomial product. It is clear $$(({\bf U}^{-1} \cdot RS(n, k))^q)^{\perp_H}=RS(n, n-k).$$ Set $$I=\{x \cdot x^j: 0 \leq j \leq k-1\}$$ and $$J=\{x^i: 0 \leq i \leq n-k-1\}.$$

We need to count how many common monomials in the set $I^q$ and $J$, where $I^q=\{{\bf e}^q: {\bf e} \in I\}$. We have $$\dim (({\bf U} \cdot RS(n, k))^q \bigcap (({\bf U} \cdot RS(n, k))^q)^{\perp_H}) =\dim (({\bf U} \cdot RS(n, k))^q \bigcap RS(n, n-k)),$$ and $$\dim (({\bf U} \cdot RS(n, k))^q \bigcap RS(n, n-k)) \geq | I^q\bigcap J|.$$ For each $0 \leq i \leq k-1$ with the expression $i=i_1q+i_2,$ where $0\leq i_1 \leq q-1$ and $0 \leq i_2 \leq q-1$, $iq=i_1q^2+i_2q$. Then $x^{iq}=x^{i_2q+i_1}$ for all elements in ${\bf F}_{q^2}$.  Since the exponents in the set $J$ are consecutive, Notice that $k \geq \frac{n}{2}$, the conclusion follows immediately.\\

It was proved that the length $q^2+1$ generalized Reed-Solomon codes over ${\bf F}_{q^2}$ of the dimension $k \geq 1+q$ can not be Hermitian self-orthogonal, see Theorem 5 in \cite{Ball1}.  However when the length $n|q^2-1$, the Hermitian hull ${\bf C} \bigcap {\bf C}^{\perp_H}$ in Theorem 2.2 is a Hermitian self-orthogonal code with big dimension. This is a generalized Reed-Solomon codes evaluated at a multiplicative subgroup.\\

{\bf Theorem 3.3.} {\em Let $n=vn_1$ be a positive integer satisfying $n_1|q^2-1$, $(q+1)|n_1$, $v \leq q-1$, and $k=k_1q+k_2$, $0 \leq k_1\leq q-1$ and $0 \leq k_2 \leq q-1$ be any given positive integer satisfying $k \geq \frac{n}{2}$. Then there exists an MDS linear code ${\bf C} \subset {\bf F}_{q^2}^n$ such that $\dim({\bf C} \bigcap {\bf C}^{\perp_H}) \geq \frac{k_1(n-k-2)}{q}$.}\\

{\bf Proof.} Let ${\bf G}$ be the subgroup of the multiplicative group of the order $n_1$. From the condition $n_1|q^2-1$ and $(q+1)|n_1$, it follows $${\bf G}\bigcap {\bf F}_q^*=\{1\}.$$ We talk $v$ cosets $b_1{\bf G}, \ldots, b_v{\bf G}$, where $b_1, \ldots, b_v $ are distinct elements in ${\bf F}_q \subset {\bf F}_{q^2}$. Set $A=\bigcup_{i=1}^v b_i {\bf G}$, then we have $$h(x)=\prod_{a \in A} (x-a)=\prod_{i=1}^v (x^{n_1}-b_i^{n_1}).$$ Then $$({\bf U}^{-1} \cdot RS(n, k))^{\perp}=RS(n, n-k),$$ where ${\bf U}=(u_1, \ldots, u_n)$, $u_i=h'(a_i)$ for $a_1, \ldots, a_n \in A$. Therefore ${\bf U}=(n_1 a_1^{n-1} B_1, \ldots, n_1a_n^{n_1-1}B_n)$, where $B_1, \ldots, B_n$ are nonzero elements in ${\bf F}_q \subset {\bf F}_{q^2}$. Set $B_i=\eta_i^{q+1}$ for some nonzero element $\eta_i \in {\bf F}_{q^2}$, and ${\bf \eta}=(\eta_1, \ldots, \eta_n)$ and ${\bf B}=(B_1, \ldots, B_n)$, we have ${\bf \eta}^{q+1}={\bf B}$.\\

We have $$({\bf B}^{-1} \cdot ({\bf \overrightarrow{h_1'}}^{-1} \cdot RS(n, k))^q)^{\perp_H}=RS(n, n-k),$$ where $h_1(x)=x^{n_1}-1$, $h_1'(x)=n_1x^{n_1-1}$, $\overrightarrow{h_1'}=(h_1'(a_1), \ldots, h_1'(a_n))$. Therefore $$(\frac{{\bf \eta}}{{\bf B}}  \cdot ({\bf \overrightarrow{h_1'}}^{-1} \cdot RS(n, k))^q)^{\perp_H}={\bf \eta}^{-q} RS(n, n-k),$$ where $\frac{{\bf \eta}}{{\bf B}}={\bf \eta}^{-q}$. We only need to calculate the dimension of $$\dim (({\bf \overrightarrow{h_1'}}^{-1} \cdot RS(n, k))^q \bigcap RS(n, n-k)).$$ Notice that $h_1(x)=n_1 x^{n_1-1}$, from a similar argument as in the proof of Theorem 3.2, the conclusion follows immediately.\\

{\bf Corollary 3.1.} {\em  Let $n=vn_1-t$ and $\frac{n}{2} \leq k=k_1q+k_2$, $0 \leq k_1 \leq q-1$, $0\leq k_2 \leq q-1$, be positive integers, where $n_1$ is a positive integer satisfying $n_1|q^2-1$ and $(q+1)|n_1$, $v\leq q-1$ and $t<n-k+1$. Then there exists an MDS linear code ${\bf C} \subset {\bf F}_{q^2}^n$ such that $\dim({\bf C} \bigcap {\bf C}^{\perp_H}) \geq \frac{k_1(n-k-2)}{q}-t$.}\\

{\bf Proof.} We consider the punctured code of a linear $[n, k]_{q^2}$ code ${\bf C} \subset {\bf F}_{q^2}^n$ to ${\bf C}_{punctured, s} \subset {\bf F}_{q^2}^{n-s}$ at the last $s$ coordinate positions. If $s< d_H({\bf C})$ the punctured mapping ${\bf C} \longrightarrow {\bf C}_{puncture, s}$ is linear and injective. Let ${\bf C}_1 \subset {\bf C}^{\perp_H}$ be the shortening code at the last $s$ coordinate positions, it is clear that ${\bf C}_{punctured, s} \subset {\bf C}_1^{\perp_H}$. On the other hand $$\dim({\bf C}_1)=n-k-s,$$ and $$\dim(({\bf C}_{punctured, s})^{\perp})=n-s-k,$$ then $$({\bf C}_{punctured, s})^{\perp_H}={\bf C}_1,$$ if $s<d_H({\bf C})$.\\

It is clear ${\bf C}_1 \subset ({\bf C}^{\perp_H})_{punctured, s}$, then $ {\bf C}_{punctured, s} \bigcap ({\bf C}_{punctured, s})^{\perp_H}$ is the shortening code of ${\bf C} \bigcap {\bf C}^{\perp_H}$. The conclusion follows immediately.\\

From this CSS construction of entanglement-assisted quantum codes, the following main results follows from the above results about the lower bound on the dimensions of Hermitian hulls.\\

{\bf Corollary 3.2.} {\em  Let $q\geq 3$ be a prime power.  Let $n$ be a positive integer satisfying $n| q^2-1$ and $k=k_q+k_2$, $0 \leq k_1\leq q-1$ and $0 \leq k_2 \leq q-1$ be any given positive integer satisfying $k \geq \frac{n}{2}$. Then there exist at least $\frac{k_1(n-k-2)}{q}+1$ MDS entanglement-assisted quantum $[[n, k-l, n-k+1, n-k-l]]_q$ codes with different nonzero $c$ parameters, where $0 \leq l \leq \frac{k_1(n-k-2)}{q}$.}\\

 {\bf Corollary 3.3.} {\em   Let $q\geq 3$ be a prime power.  Let $n=vn_1$ be a positive integer satisfying $n_1|q^2-1$, $(q+1)|n_1$, $v\leq q-1$, and $k=k_1q+k_2$, $0 \leq k_1\leq q-1$ and $0 \leq k_2 \leq q-1$ be any given positive integer satisfying $k \geq \frac{n}{2}$. Then there exist at least $\frac{k_1(n-k-2)}{q}+1$ MDS entanglement-assisted quantum $[[n, k-l, n-k+1, n-k-l]]_q$ codes with different nonzero $c$ parameters, where $0 \leq l \leq \frac{k_1(n-k-2)}{q}$.}\\

 {\bf Corollary 3.4.} {\em  Let $q\geq 3$ be a prime power.  Let $n=vn_1-t$ and $\frac{n}{2} \leq k=k_1q+k_2$, $0 \leq k_1 \leq q-1$, $0\leq k_2 \leq q-1$, be positive integers, where $n_1$ is a positive integer satisfying $n_1|q^2-1$ and $(q+1)|n_1$, $v\leq q-1$ and $t<n-k+1$. Then there exists at least $\frac{k_1(n-k-2)}{q}+1-t$ MDS entanglement-assisted quantum $[[n, k, d, c]]_q$ codes with different nonzero $c$ parameters.}\\

 Since $n=q+1$ is a factor of $q^2-1$, we can take $n_1=q+1$ and $v=1, 2, \ldots, q-1$ and $t < \min\{n-k+1, q\}$ in Corollary 2.4. Then for the given length $O(q^2)=n \leq q^2+1$ and the given distance $O(q^2)=d \leq \frac{n+2}{2}$, there are at least $O(q^2)$ different MDS entanglement-assisted quantum codes with different $c$ parameters.\\

\section{Hermitian hulls of generalized twisted Reed-Solomon codes}

We consider a class of twisted Reed-Solomon codes introduced in \cite{BPR}. For a twisted Reed-Solomon code ${\bf C}_{{\bf \alpha}, {\bf t}, {\bf h}, {\bf \eta}}^{n, k}$ evaluated at a subgroup of the multiplicative group ${\bf F}_q^*$, the dual is equivalent to another twisted Reed-Solomon code ${\bf C}_{{\bf \alpha}, k-{\bf h}, n-k-{\bf t}, -{\bf \eta}}^{n, n-k}$. If $q=2^s$, then for suitable ${\bf h}, {\bf t}$ and ${\bf \eta}$, from the method proposed in \cite{HChen22}, it is easy to get a self-dual twisted Reed-Solomon code. Some of them are MDS codes.\\

In this section we calculate the Hermitian hulls of the generalized twisted Reed-Solomon codes arising from the subgroup of  f${\bf F}_{q^2}^*$, see \cite{BPR}. Let $n$ and $k$ be two positive integers satisfying $n|q^2-1$ and $k \leq n-1$. Let $\alpha_1, \ldots, \alpha_n$ be $n$ distinct elements in the finite field ${\bf F}_{q^2}$ such that ${\bf \alpha}=\{\alpha_1, \ldots, \alpha_n\}$ is a multiplicative subgroup of ${\bf F}_{q^2}^*$. Let $\eta$ be a nonzero element of ${\bf F}_{q^2}$. Set ${\bf g}_0=1+\eta x^k$, ${\bf g}_1=x$, \ldots, ${\bf g}_{k-1}=x^{k-1}$. Let ${\bf P}(\eta, k)$ be the linear span over ${\bf F}_{q^2}$ by ${\bf g}_0, \ldots, {\bf g}_{k-1}$. The linear code ${\bf C}_{{\bf \alpha}, \eta, k}$ is the evaluation code of these polynomials in ${\bf P}(\eta, k)$ at the above $n$ distinct elements of the subgroup ${\bf \alpha}$. The dimension of the Schur square of ${\bf C}_{{\bf \alpha}, \eta, k}$ is at least $2k$, see \cite{BPR}. Thus this code is not equivalent to a Reed-Solomon code when $2k \leq n$, even when it is MDS.\\

Set ${\bf h}_0=1$, ${\bf h}_1=x$, \ldots, ${\bf h}_{n-k-1}=x^{n-k-1}-\eta x^{n-1}$. Let ${\bf P}(-\eta, n-k)^{\perp}$ be the linear span over ${\bf F}_{q^2}$ by ${\bf h}_0, \ldots, {\bf h}_{n-k-1}$. Let ${\bf C}_{{\bf \alpha}, -\eta, n-k}$ be the linear code which is the evaluation code of these polynomials in ${\bf P}(-\eta, n-k)^{\perp}$ at the above $n$ distinct elements of the subgroup ${\bf \alpha}$.\\

{\bf Proposition 4.1 (or see Theorem 5 in \cite{BPR}).} {\em The dual code of ${\bf C}_{{\bf \alpha}, \eta, k}$ is of the form ${\bf U}^{-1} \cdot {\bf C}_{{\bf \alpha}, -\eta, n-k}$, where ${\bf U} \in {\bf F}_{q^2}^n$ is a Hamming weight $n$ vector as in Theorem 3.2.}\\

{\bf Proof.} First of all ${\bf C}_{{\bf \alpha}, \eta, k} \subset RS(n, k+1)$, then ${\bf U}^{-1} \cdot RS(n, n-k-1) \subset {\bf C}_{{\bf \alpha}, \eta, k}^{\perp}$. We only need to prove that the products of ${\bf h}_{n-k-1}$ and ${\bf g}_0, \ldots, {\bf g}_{k-1}$ contain no terms $x^{v}$ satisfying $v \geq n-1$. Because ${\bf \alpha}=\{\alpha_1, \alpha_2, \ldots, \alpha_n\}$ is a multiplicative subgroup of ${\bf F}_{q^2}^*$, then $x^n=1$ for each element in this subgroup. Hence $(1+\eta x^k)(x^{n-k-1}-\eta x^{n-1})=x^{n-k-1}-\eta^2 x^{k-1}$, and $h_{n-k-1}x^i=x^{n-k-1+i}-\eta x^{i-1}$ for $i=1, \ldots, k-1$, when evaluated at elements of this subgroup ${\bf \alpha}=\{\alpha_1, \ldots, \alpha_n\}$. The conclusion follows immediately.\\

{\bf Proposition 4.2 (or see Theorem 1 of \cite{BPR}).} {\em If $\eta$ is not in the subgroup ${\bf \alpha}$, then ${\bf C}_{{\bf \alpha}, \eta, k}$ is an MDS code.}\\

{\bf Theorem 4.1.} {\em Let $n$ be a divisor of $q^2-1$ as in Theorem 3.2 and $k$ be a positive integer satisfying $\frac{n}{2} \leq k \leq n$. The dimension of the Hermitian hull of the code $({\bf U}^{-1} \cdot {\bf C}_{{\bf \alpha}, \eta, k})^q$ is at least $\frac{k(n-k-2)}{q^2}$.}\\

{\bf Proof}. It is clear that the code ${\bf U}^{-1} \cdot {\bf C}_{{\bf \alpha}, \eta, k}$ is the evaluation code of linear combinations of $x{\bf g}_0, \ldots, x{\bf g}_{k-1}$. We have $(({\bf U}^{-1} \cdot {\bf C}_{{\bf \alpha}, \eta, k})^q)^{\perp_H}={\bf C}_{{\bf \alpha}, -\eta, n-k}$, which is the evaluation code of ${\bf h}_0, \ldots, {\bf h}_{n-k-1}$. Set $I=\{x x^j : 1 \leq j \leq k\}$ and $J=\{x^j: 1\leq i \leq n-k-2\}$. We need to count how many common monomials in $I^q \bigcap J$. The conclusion follows from a similar argument as the proof of Theorem 3.2.\\

From Theorem 4.2 it is proved that MDS entanglement assisted quantum codes can also be constructed from the generalized twisted Reed-Solomon codes.\\

\section{Conclusion}
Construction of MDS quantum codes and MDS entanglement-assisted quantum codes have been a long-time challenging problem in the theory of quantum error correction codes. Presently only few MDS quantum codes have been constructed for restricted lengths. Before our paper \cite{HChen221}, MDS entanglement-assisted quantum codes have only been constructed for very specific lengths and distances. In this paper we showed that for each possible length and each possible distance there are at least one MDS entanglement-assisted quantum code. On the other hand when the length $n$ and the distance are of the form $O(q^2)$, there exist at least $O(q^2)$ different MDS entanglement-assisted quantum codes. Therefore there are much more MDS entanglement-assisted quantum codes than MDS quantum codes. This is natural from the physical point of view. Our method can be applied to construct MDS entanglement-assisted quantum codes from the generalized MDS twisted Reed-Solomon codes. Therefore many new MDS entanglement assisted quantum codes from the MDS twisted Reed-Solomon codes were constructed.\\


\begin{thebibliography}{10}




\bibitem{AKS} S. A. Aly, A. Klappenecker and P. K. Sarvepalli, On quantum and classical BCH codes, IEEE Trans. Inf. Theory, vol. 53, no. 3, pp. 1183-1188, 2007.

\bibitem{Ball} S. Ball, On large subsets of a finite vector space in which every subset of basis size is a basis, J. EMS, vol. 14, pp. 733-748, 2012.

\bibitem{Ball1} S. Ball, Some constructions of quantum MDS codes, Des., Codes and Cryptogra. vol. 89, pp. 811-821, 2021.


\bibitem{Ball2} S. Ball and R. Vilar, Determining when a truncated generalised Reed-Solomon code is Hermitian self-orthogonal, IEEE Trans. Inf. Theory, vol. 68, pp. 3796-3805,  2022.

\bibitem{Ball3} S. Ball and R. Vilar, The geometry of Hermitian orthogonal codes, Journal of Geometry, vol. 113, artical no. 7, 2022.

\bibitem{Ball4} S. Ball, A. Centelles and F. Huber, Quantum error-correcting codes and their geometries, Annale de l'institut Henri Poincare, to appear, 2022.


\bibitem{Barta} M. Berta, H. Gharibyan and M. Walter, Entanglement-assisted capacities of compound quantum channels, IEEE Trans. Inf. Theory, vol. 63, no. 5, pp. 3306-3321, 2017.

\bibitem{BPR} P. Beelen, S. Puchinger and J. Rosenkilde, Twisted Reed-Solomon codes, IEEE Trans. Inf. Theory, vol. 68, no. 5, pp. 3047-3061, 2022.

\bibitem{Brun} T. Brun, I. Devetak and Min-Hsiu Hsieh, Correcting quantum errors with entanglemnent, Science, vol. 304 (5798), no.6, pp. 436-439, 2006.

\bibitem{BDH} T. Brun, I. Devetak and Min-Hsiu Hsieh, Catalytic quantum error correction, IEEE Trans. Inf. Theory, vol. 60, pp. 3073-3089, 2014.

\bibitem{CShor} A. Calderbank and P. W. Shor, Good quantum error-correcting codes exist, Phys. Rev. A, Gen. Phys., vol. 54, no. 2, pp. 1098-1105, 1996.

\bibitem{CRSS} A. Calderbank, E. M. Rains, P. W. Shor and N. J. A. Sloane, Quantum error correction via codes over $GF(4)$, IEEE Trans. Inf. Theory, vol. 44, no. 4, pp. 1369-1387, 1998.



\bibitem{HChen1} H. Chen, Some good quantum error-correcting codes from algebraic geometric codes, IEEE Trans. Inf. Theory, vol. 47, no. 5, pp. 2059-2061, 2001.


\bibitem{HChen2} H. Chen, C. Xing and S. Ling, Asymptotically good quantum codes exceeding the Ashikhmin-Litsyn-Tsafasman bound,  IEEE Trans. Inf. Theory, vol. 47, no. 5, pp. 2055-2058, 2001.


\bibitem{HChen3} H. Chen, C. Xing and S. Ling, Quantum codes from concatenated algebraic geometric codes, IEEE Trans. Inf. Theory, vol. 51, no. 8, pp. 2915-2920, 2005.



\bibitem{HChen22} H. Chen, On hull-variation problem of equivalent linear codes, arXiv:2206.14516, 2022.

\bibitem{HChen221} H. Chen, New MDS entanglement-assisted quantum codes from Hermitian self-orthogonal codes, arXiv:2206.13995, 2022.

\bibitem{CZJL} X. Chen, S. Zhu, W. Jiang and G. Luo, A new family of EAQMDS codes constructed from constacyclic codes, Des., Codes and Cryptogra., vol. 89, pp. 2179-2193, 2021.

\bibitem{CZJ} X. Chen, S. Zhu and W. Jiang, Cyclic codes and some new entanglement-assisted quantum MDS codes, Des., Codes and Cryptogra., vol. 89, pp. 2533-2551, 2021.

\bibitem{CMPR} R. B. Christensen, C. Munuera, F. R. F. Pereira and D. Ruano, An algorithmic approach to entanglement-assisted quantum error-correcting codes from Hermitian curves,  Adv. Math. Commun., online first version, 2022.
    
    
\bibitem{FanPoor} J. Fan, J. Li, Y. Zhou and H. V. Poor, Entanglement-assisted quantum concatenated codes, Proc. NAS, DOI.ORG/10/1073/PNAS/220235119, June, 2022.


\bibitem{FFLZ20} W. Fang, F. Fu, L. Li and S. Zhu, Euclid and Hermitian hulls of MDS codes and their application to quantum codes, IEEE Trans. Inf. Theory vol. 66. no. 6, pp. 3527-3537, 2020.



\bibitem{Fattal} D. Fattal, T. S. Cubitt, Y. Yamamoto, S. Bravyi and I. L. Chuang, Entanglement in the stablizer formalism, arXiv:quant-ph/04-6168, 2004.

\bibitem{Fujiwara} Y. Fujiwara, D. Clark, P. Vandendriessche, M. De Boeck and V. D. Tonchev, Entanglement-assisted quantum low-density parity check code, Phys. Rev. A, vol. 88, 012318, 2013.





\bibitem{GYHZ} Y. Gao, Q. Yue, X. Huang and J. Zeng, Hulls of generalized Reed-Solomon codes via Goppa codes and their applications to quamtum codes, IEEE Trans. Inf. Theory, vol. 67, no. 10, pp. 6619-6626, 2021.

\bibitem{GMCR} C. Galindo, F. Hernando and D. Ruano, Entanglement-assisted quantum codes from RS codes and BCH codes with extension degree two, Quantum Inf. Process., vol. 20, pp. 158, 2016.



\bibitem{GuoLi} G. Guo and R. Li, Hermitian self-dual GRS and entended GRS codes, IEEE Commun. Lett., vol. 25, no. 4, pp. 1062-1065, 2021.




\bibitem{GG08} M. Grassl and T. A. Gulliver, On self-dual MDS codes, Proc. Int. Symp. Inf. Theory, pp. 1954-1957, 2008.

\bibitem{GHW} M. Grassl, F. Huber and A, Winiter, Entropic proofs of Singleton bounds for quantum error-correcting codes, IEEE Trans. Inf. Theory, vol. 68, no. 6, pp. 3942-3950, 2021.

\bibitem{Guar} G. G. La Guardia, New quantum MDS codes, IEEE Trans. Inf. Theory, vol. 57, no. 8, pp. 5551-5554, 2011.



\bibitem{HCX} X. He, L. Xu and H. Chen, New $q$-ary quamtum MDS codes with distances bigger than $\frac{q}{2}$, Quantum Inf. Process., vol. 15, pp. 2745-2758, 2016.

\bibitem{HDB} M-H. Hsieh, I Devetak and T. A. Brun, General entanglement-assisted quantum error-correcting codes, Phys. Rev. A, vol. 76, 062313, 2007.

\bibitem{HBD} M-H. Hsieh, T. A. Brun and I Devetak, Entanglement-assisted quantum quasi-cyclic low-density parity-check codes, Phys. Rev. A, vol. 79, 032340, 2009.

\bibitem{HP} W. C. Huffman and V. Pless, Fundamentals of error-correcting codes, Cambridge University Press, Cambridge, U. K., 2003.


\bibitem{JX} L. Jin and C. Xing, Euclid and Hermitian self-orthogonal algebraic geometric codes and their applications to quantum codes, IEEE Trans. Inf. Theory, vol. 58, no. 8, pp. 5484-5489, 2012.





\bibitem{JX17}  L. Jin and C. Xing, New MDS self-dual codes from generalized Reed-Solomon codes,  IEEE Trans. Inf. Theory, vol. 63, no. 3, pp. 14-34-1438, 2017.

\bibitem{LBW} C-Y. Lai, T. A. Brun and M. M. Wilde, Duality in entanglement-assisted quantum codes, IEEE Trans. Inf. Theory, vol. 59, no. 6, pp. 4020-4024, 2013.



\bibitem{LCC} G. Luo, X. Cao and X. Chen, MDS codes with hulls of arbitray dimensions and their quantum error correction, IEEE Trans. Inf. Theory, vol. 65, no. 5, pp. 2944-2952, 2019.

\bibitem{LES22} G. Luo, M. F. Ezerman and S. Ling, Entanglement-assisted and subsystem quantum codes: new propagation rule and construction, arXiv:2206.09782, 2022.

\bibitem{LEGL22} G. Luo, M. F. Ezerman, M. Grassl and S. Ling, How much entanglement does a quantum code need? arXiv4395197, 2022.


\bibitem{KZ13} X. Kai and S. Zhu, New quantum MDS codes from negacyclic codes, IEEE Trans. Inf. Theory, vol. 59, no. 2, pp. 1193-1197, 2013.


\bibitem{KZL14} X. Kai, S. Zhu and P. Li, Constacyclic codes and some new quantum MDS codes, IEEE Trans. Inf. Theory, vol. 60, no. 4, pp. 2080-2086, 2014.

\bibitem{KS} A. Klappenecker and P. K. Sarvepalli, Clifford code constructions of operators quantum error correcting codes, IEEE Trans. Inf. Theory, vol. 54, no. 12, pp. 5760-5765, 2008.

\bibitem{Kor} M. E. Koroglu, New entanglement-assisted MDS quantum codes from constacyclic codes, Quantum Inf. Process., vol. 18, pp. 1-18, 2018.



\bibitem{KHB} I. Kremsky, M-H. Hsieh and T. A. Brun, Classical enhansement of quantum error-correcting, Phys. Rev. A, vol. 78, 012341, 2008.


\bibitem{Pellikaan} F. R. F. Pereira, R. Pellikaan, G. G. La Guardia and F. Marcos, Entanglement-assisted quantum codes from algebraic geometry codes, IEEE Trans. Inf. Theory, vol. 67, no. 11, pp. 7110-7120, 2021.




\bibitem{Shor} P. W. Shor, Scheme for redcuing decoherence in quantum memory, Phys. Rev. A, vol. 52, pp. R2493-2496, 1995.

\bibitem{Steane1} A. M. Steane, Error-correcting codes in quantum theory, Phys. Rev. Lett., vol. 77, no. 5, pp. 793-797, 1996.




\bibitem{Steane} A. M. Steane, Mutiple particle interference and quantum error correction, Proc. Roy. Soc. London, vol. 452, pp. 2551-2577, 1996.



\bibitem{WKB} M. M. Wilde, H. Krovi and T. A. Brun, Entanglement-assisted quantum error correction with linear optic, Phys. Rev. A, vol. 76, 052308, 2007.

\bibitem{WB} M. M. Wilde, and T. A. Brun, Optimal entanglement formulae for entanglement-assisted quantum coding, Phys. Rev. A, vol. 77, 064302, 2008.


\end{thebibliography}
\end{document}